%% file: main.tex
\documentclass[10pt, conference]{IEEEtran}
\IEEEoverridecommandlockouts
\usepackage{cite}
\usepackage{amsmath,amssymb,amsfonts}
\usepackage[ruled,vlined]{algorithm2e}
\LinesNumbered

\SetCommentSty{mycommfont}

\makeatletter
\renewcommand{\SetKwInOut}[2]{
  \sbox\algocf@inoutbox{\KwSty{#2\algocf@typo:}}
  \expandafter\ifx\csname InOutSizeDefined\endcsname\relax
    \newcommand\InOutSizeDefined{}\setlength{\inoutsize}{\wd\algocf@inoutbox}%
    \sbox\algocf@inoutbox{\parbox[t]{\inoutsize}{\KwSty{#2\algocf@typo\hfill:}}~}\setlength{\inoutindent}{0em}
  \else
    \ifdim\wd\algocf@inoutbox>\inoutsize
    \setlength{\inoutsize}{\wd\algocf@inoutbox}
    \sbox\algocf@inoutbox{\parbox[t]{\inoutsize}{\KwSty{#2\algocf@typo\hfill:}}~}\setlength{\inoutindent}{0em}
    \fi
  \fi
  \algocf@newcommand{#1}[1]{
    \ifthenelse{\boolean{algocf@hanginginout}}{\relax}{\algocf@seteveryparhanging{\relax}}
    \ifthenelse{\boolean{algocf@inoutnumbered}}{\relax}{\algocf@seteveryparnl{\relax}}
    {\let\\\algocf@newinout\hangindent=\inoutindent\hangafter=1\parbox[t]{\inoutsize}{\KwSty{#2\algocf@typo\hfill:}}~##1\par}
    \algocf@linesnumbered
    \ifthenelse{\boolean{algocf@hanginginout}}{\relax}{\algocf@reseteveryparhanging}
  }}
\makeatother

\usepackage{graphicx}
\usepackage{textcomp}
\usepackage{xcolor}

\usepackage{multicol}
\usepackage{graphicx}
\usepackage{multirow}
\usepackage{float}
\usepackage{paralist}

\usepackage{xcolor}
\usepackage{amsmath}
\usepackage{amsfonts}
\usepackage{subcaption}
\usepackage{comment}
\usepackage{todonotes}

\usepackage{mwe}
\usepackage{caption}
\usepackage{graphicx}

\usepackage{subcaption}
\usepackage{svg}
\usepackage[shortlabels]{enumitem}
\usepackage{siunitx}
\usepackage{wrapfig}
\usepackage{amssymb}
\usepackage{adjustbox}
\usepackage{booktabs} 

\usepackage{makecell}
\usepackage{xcolor,pifont}

\setlist{nolistsep}

\newcommand{\mand}{\mbox{$~\wedge~$}}
\newcommand{\snn}{\mbox{$\mathcal{N}$}}
\newcommand{\ann}{\mbox{$\mathcal{A}$}}
\newcommand{\R}{\mbox{$~\mathbb{R}~$}}

\newtheorem{definition}{Definition}%[section]

\usepackage{soul}

\def\BibTeX{{\rm B\kern-.05em{\sc i\kern-.025em b}\kern-.08em
    T\kern-.1667em\lower.7ex\hbox{E}\kern-.125emX}}
\begin{document}

\title{
% Safety Verification of Spiking Neural Network Controllers\\
%Verifying Safe Range Specifications of Spiking Neural Network Controllers
Configuring Safe Spiking Neural Controllers for Cyber-Physical Systems through Formal Verification
}

% \and
% \and
% \and
% \and
% \and
\author{Arkaprava Gupta,
		Sumana Ghosh, %\orcid{0000-0002-5999-3313}
            Ansuman Banerjee, %{}
		and Swarup Kumar Mohalik
  %\orcid{}
        \thanks{Arkaprava Gupta (email: arkaprava.gupta@gmail.com) and Swarup Kumar Mohalik (email: swarup.kumar.mohalik@ericsson.com) are with Ericsson India Pvt Ltd,  Bangalore 560048, India.
  %is with Kalinga Institute of Industrial Technology, Odisha 751024, India,  
  Sumana Ghosh (email: sumana@isical.ac.in) and Ansuman Banerjee (ansuman@isical.ac.in) are with Indian Statistical Institute, Kolkata, India. This work was partially supported by a funding from Ericsson India Pvt Ltd.}
    %\thanks{This work was partially supported by a funding from Ericsson India Pvt Ltd.}
  }
  
\maketitle

\input{sections/abstract}


\input{sections/intro}
\input{sections/preliminaries}
\input{sections/formal_model}
\input{sections/solutiondetails}
\input{sections/experiment}

\input{sections/results}
\input{sections/related}
\input{sections/conclusion}

% \newpage
%\clearpage
\bibliographystyle{unsrt}

\bibliography{main}
\end{document}

%% file: sections/abstract.tex
\begin{abstract}

Spiking Neural Networks (SNNs) are a subclass of neuromorphic 
models that have great potential to be used as controllers in Cyber-Physical Systems (CPSs) due to their energy efficiency. They can benefit from the prevalent approach of first training an Artificial Neural Network (ANN) and then translating to an SNN with subsequent hyperparameter tuning.
The tuning is required to ensure that the resulting SNN is accurate
with respect to the ANN in terms of metrics like Mean Squared Error (MSE).
However, SNN controllers for safety-critical CPSs must also satisfy 
safety specifications, which are not guaranteed by the conversion
approach. In this paper, we propose a solution which tunes the
{\em temporal window} hyperparameter of the translated SNN to ensure 
both accuracy and compliance with the safe range specification 
that requires the SNN outputs to remain within a safe range. 
The core verification problem is modelled using mixed-integer
linear programming (MILP) and is solved with Gurobi.
When the controller fails to meet the range specification, we compute 
tight bounds on the SNN outputs as feedback for the CPS developer. 
To mitigate the high computational cost of verification, we integrate
data-driven steps to minimize verification calls. Our approach
provides designers with the confidence to safely integrate energy-efficient 
SNN controllers into modern CPSs. We demonstrate our
approach with experimental results on five different benchmark
neural controllers.
\end{abstract}

%% file: sections/intro.tex
%-------------------------------------------
% \textcolor{blue}{Reviewer's comments remaining to be addressed:}
% \begin{enumerate}
    
% \item \textcolor{blue}{ Define amplitude formally. What if no spike? Is negative spikes possible?
% \item ANN to SNN conversion to be explained.
% \item Provide time complexity of the overall algorithm.
% \item Formal definition for SNN and SNN-controlled systems 
% \item Is considering 1\% a typical approach? Please provide a reference.}
% \end{enumerate}
%-----------------------------------------

\section{Introduction} \label{sec:intro}
\begin{figure}[h]
     \centering
     \scalebox{0.8}{
     \includegraphics[width=1\linewidth]{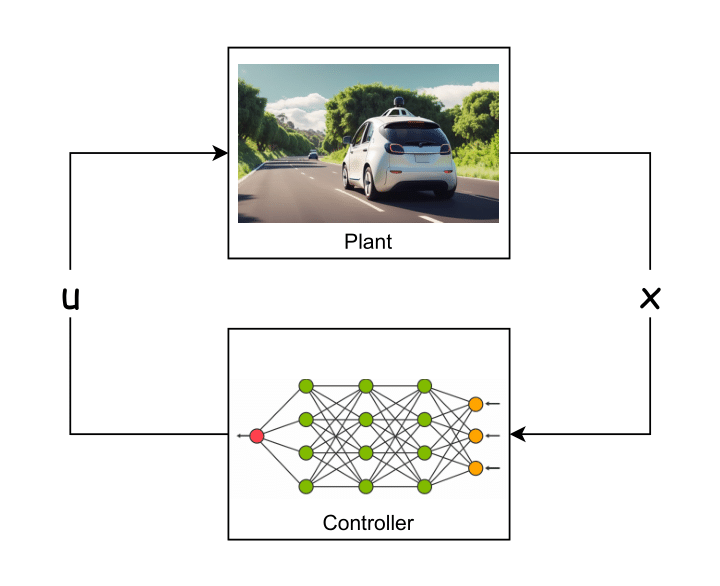}
     }
     \caption{NN-Controlled CPS}
     \label{fig:cps2}
 \end{figure}

\noindent
Cyber-physical systems (CPSs) are engineered systems created through the seamless integration of computational algorithms and physical components. A typical CPS comprises a physical system (plant) and a controller that monitors the plant's behavior and provides control inputs to steer the system towards specified goals. As physical systems grow more complex and demand novel functionalities, traditional control design methodologies are becoming inadequate. Consequently, these methods are increasingly being replaced by learning-based models such as artificial neural networks (ANNs)~\cite{Mnih2015HumanlevelCT, Pavithra, polar}. Over recent years, ANNs have been applied to various autonomous safety-critical systems, from aerospace to unmanned vehicles~\cite{Bojarski2016EndTE, Julian2018DeepNN, Katz2017ReluplexAE}. Figure~\ref{fig:cps2} shows an example of an ANN-controlled CPS.

A Spiking Neural Network (SNN) is a special type of neural network that can achieve accuracies comparable to traditional ANNs, with the additional advantage of lower energy requirements~\cite{main, vmcai, partharoop2023, snnconvwang}. Recent times have seen tremendous growth in neuromorphic hardware~\cite{neuromorphic_computing, neuromorphic}, SNN models and their usage in many real-world applications including some safety-critical CPSs~\cite{main, partharoop2023}. Neuromorphic hardware is a type of computer architecture that mimics the brain's neural networks to process information in a brain-like manner, using artificial neurons and synapses.
The shift from ANN-based control to SNN-based alternatives has accelerated recently due to increasing energy concerns and rising demand for power in CPS applications.

SNNs, similar to ANNs, utilize various neuron types, prominent among them being the Leaky-Integrate-Fire (LIF) neuron and the Izhikevich neuron. SNNs with Spiking Rectified Linear Activation (SRLA)~\cite{Bekolay2014} have been gaining attention recently~\cite{asyncSN} because of their seamless conversion from ANNs with ReLU activation~\cite{Bekolay2014}, thereby, making them efficacious for both classification and regression problems. Moreover, ANN-based controllers mostly use ReLU
activations because of their robust performance in noisy environments and with complex nonlinear plant dynamics~\cite{polar, Katz2017ReluplexAE, dutta2017output,ARCH19, ARCH22}. This motivates our choice of ReLU-like activation for SNNs, in particular, SNNs with SRLA in this study.

An increasingly popular trend in the SNN community today~\cite{snnconvwang,snnconvcao} is to first train an ANN model and then transform it to an SNN ensuring a high level of accuracy relative to the original ANN in terms of metrics like Mean Squared Error (MSE).
This is being done primarily to avoid the resource-hungry and complex training algorithms for SNNs, and also to take advantage of the already tested and/or verified ANN models. 
During this conversion, a crucial hyperparameter to be fixed is the {\em temporal window} or the input sequence length (called as NUMSTEPS), of the SNN. Larger temporal windows can improve accuracy but increase computational demands and latency. 

State-of-the-art ANN-SNN conversion algorithms attempt to strike a balance between accuracy and latency during the selection of NUMSTEPS, taking into consideration constraints such as {\em the SNN latency must be accommodated within the control period}. However, they fall short when this conversion is done to obtain an SNN controller for CPSs because they do not provide explicit guarantees for the safety requirements in safety-critical applications. %such as autonomous vehicles, robotics, and medical devices.
The safety guarantees may be provided in two ways: (1) by establishing an equivalence modulo the safety requirements between the ANN and the corresponding SNN or (2) by verifying the SNN against the safety requirements. While there has been a handful of research efforts in establishing the equivalence between ANN and SNN~\cite{snnconvwang, snnconvsengupta,snnconvgao} (though not in the context of CPS applications), and a plethora of work in the verification of ANN-controlled CPSs~\cite{polar, hoang_19, reachSherlock, 2020reachnn, verisig2}, the problem of safety assurance of the translated SNN controllers remains completely unexplored in literature. In this work, we primarily address this problem, illustrating the solution for the safe range requirement which ensures that the output of the SNN always remains within a safe range. There have been significant efforts in MILP-based modeling of neural networks. Our work is an extension of the same, but in the context of SNNs along the lines of~\cite{vmcai}. The differentiating factor with respect to ANN encodings is the way spikes are modelled and the execution semantics of neurons in SNNs. The FV problem for SNNs is hard in general \cite{vmcai}, as in the case of ANNs \cite{Katz2017ReluplexAE}.

In particular, we show that the determination of the temporal window (NUMSTEPS) for the SNN, aimed at achieving the accuracy objective, is inadequate for guaranteeing the safe range requirement. We then provide a solution to find the least value of NUMSTEPS for the generated SNN such that it achieves both the accuracy objective and is compliant with the safe range requirement. We believe that our solution will provide designers with the confidence to safely integrate energy-efficient SNN controllers into modern CPSs.  
To the best of our knowledge, this is the first proposal in the literature that addresses the design of SNN controllers ensuring their given safe range specifications.
The specific contributions of the paper are outlined in the following:
\begin{enumerate}
    \item we derive an upper bound for NUMSTEPS from the control period of the CPS and the execution time for one step of the SNN, and provide an iterative procedure to compute the least value of NUMSTEPS ensuring required accuracy and safe range compliance,
    \item given an SNN, a {\em fixed} value of NUMSTEPS, and a specified range for SNN outputs, we provide a formal verification procedure by encoding the range check specification and SNN semantics using Mixed-Integer Linear Programming (MILP) and solving it using Gurobi~\cite{gurobi}, 
    \item we give an iterative bound tightening procedure with formal verification as a subroutine to produce tight bounds for the output ranges of the SNN for which the given safe range specification is not satisfied,
    \item since formal verification can be costly in terms of computation time, we provide a method that tries to minimize the verification steps by incorporating data-driven accuracy calculation and range violation checks which can be done quickly,
    
    \item we validate our framework by conducting experiments on five benchmark neural network controllers publicly available 
    in~\cite{ARCH19,ARCH22}.
\end{enumerate}
\noindent
The rest of this paper is organized as follows. Section~\ref{sec:preliminaries} presents a background on ANN, SNN and ANN-SNN conversion. Section~\ref{sec:formal_model} discusses our 
system model and the problem definition with a motivating example.  We present the solution outline followed by details of the methodology in Section~\ref{sec:detailed_methodology}. Implementation details, experiments and analysis are presented in Section~\ref{sec:implementation}. Section~\ref{sec:related} points to the related work. Section~\ref{sec:conclusion} concludes the paper with a summary and future plans.

%% file: sections/preliminaries.tex
\section{Preliminaries}
\label{sec:preliminaries}
\noindent
\subsection{Artificial Neural Network (ANN) Controller}
\noindent
An ANN-based controller for a CPS is trained using carefully constructed datasets including plant observables as input features and control input for the plant as its output features. The ANN is tested using datasets that should capture various requirements of the CPS. A critical requirement is that of  {\em safe range} which mandates that each output of the ANN (control input to the plant) is bounded by a given range that is dependent upon the plant state.   
Note that the derivation can be done employing optimization\cite{Pavithra} or formal verification techniques\cite{xiang2018verification}.

\subsection{Spiking Neural Network (SNN) Controller } 
\label{sec:background}
\noindent
An SNN consists of neurons that process a sequence of spike inputs over the temporal window (i.e., NUMSTEPS) and produce spike outputs of the same length. Depending upon the nature of neurons, these spikes can have binary or non-binary amplitudes. The final output is an aggregation of the output spikes (e.g., the average of the spike values over NUMSTEPS). We denote by $\mathcal{N}_T$ an SNN $\mathcal{N}$ where NUMSTEPS has been set to $T$.  
 We make the following assumption about the SNN when they are used as controllers for CPSs. First, when the plant observables $I$ are sent to the SNN $\mathcal{N}_T$, the input $I$ is repeated for all the $T$ steps as input. For all practical purposes, we assume that all the inputs are bounded. Second, there is an upper bound $T_{up}$ on $T$, computed from the control period $p$ and the execution time $e$ for one step of the SNN: $T \leq T_{up} = \lfloor{p/e}\rfloor$.
 
 Unlike ANNs, the neurons of an SNN store the potential generated as a result of the input synapses and the connection weights modulating them. This is known as {\em the membrane potential}. Once a given potential {\em threshold} is crossed, the neuron fires a spike and the stored potential is reset for a refractory period - this limits a neuron's firing frequency.

\textbf{SNN with Spiking Rectified Linear Activation (SRLA)}: A variety of spiking neuron models exist exhibiting tradeoffs between biological accuracy and computational feasibility. As mentioned earlier, in this work, we consider neurons with SRLA. SRLA neurons accumulate potential ($P$) by calculating the product ($\Delta$) of the incoming weights ($w$) with the output spike amplitudes of the connecting neurons in the previous layer until they reach their threshold ($\theta$). At every time step, the existing membrane potential of a neuron leaks and only a portion of it is retained. This is determined by the leak factor, denoted by $\lambda \in \R$. Upon reaching the threshold, the neurons spike with an amplitude
$\lfloor P/\theta\rfloor$, and their membrane potential is reset to $P - \lfloor P/\theta\rfloor$. 
In an SNN with SRLAs, all the internal nodes have SRLA activation, while the output is just an average over the linear sum of the instant potentials.
An example SNN with the SRLA neurons is given below with a short explanation of its working.

\begin{figure}[h]
    \centering
    \includegraphics[width=0.7\linewidth]{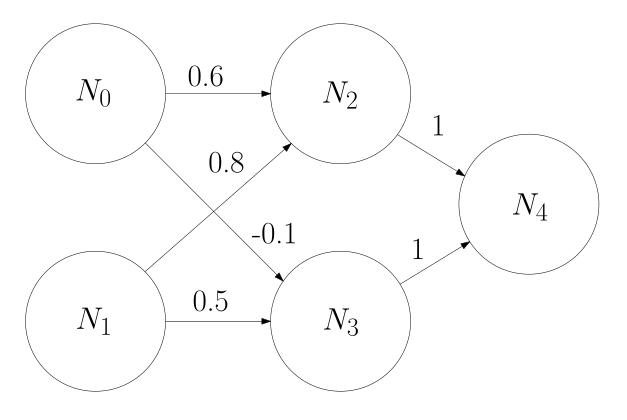}
    \caption{An Example SNN } 
    \label{fig:enter-label}
\end{figure}

\noindent
{\bf Example:} Consider the SNN shown in Figure~\ref{fig:enter-label} where NUMSTEPS is set to 6. 

Each neuron has its initial stored potential as $0$ and the threshold as $1$. 
In this example, we consider the leak factor as 1 (no leakage) for all neurons. The SNN receives an input of $5$ and $2$ through its two input neurons $N_0$ and $N_1$ respectively over $6$
timesteps, i.e. on inputs $N_0$ and $N_1$, the vector 
$\langle (5,2), (5,2), (5,2), (5,2), (5,2), (5,2) \rangle$ is received. The computations in the SNN for the first step are as follows:
 the input neurons $N_0$ and $N_1$ spike with amplitudes $5 (\lfloor\frac{5}{1} \rfloor$) and $2(\lfloor\frac{2}{1} \rfloor$) respectively.
The stored potential of neuron $N_2$ at time step $1$ is updated to $4.6 = 0.6 \times 5 + 0.8 \times 2$, since the weights of the input synapses $N_0$\_$N_2$ and $N_1$\_$N_2$ are $0.6$ and $0.8$ respectively. $N_2$ spikes because the stored potential is now greater than the threshold 1. It spikes with amplitude 4($\lfloor\frac{4.6}{1} \rfloor$), resetting the stored potential to 0.6. 
Similarly, the stored potential $N_3$ is updated to a value of $0.5 = -0.1 \times 5 + 0.5 \times 2$. There is no spike output from $N_3$ because the stored potential is less than the threshold 1.
Finally, $N_4$ outputs a value of $4$ ($= 1 \times 4$~ (spike output of $N_2$) + $1 \times 0$~ (spike output of $N_3$)).

\begin{table}[htb!]
    \centering
    \renewcommand{\arraystretch}{1.1}
\begin{tabular}{@{}c|ccc|ccc@{}}
\toprule

\multirow{2}{*}{{{\bf Time}}} & \multicolumn{3}{c|}{{{\bf Stored Potential}}} & \multicolumn{3}{c}{{{\bf Spike Outputs}}} \\ 
\cmidrule{2-7}
&
\multicolumn{1}{c}{$N_2$} & \multicolumn{1}{c}{$N_3$} & $N_4$  & \multicolumn{1}{c}{$N_2$} & \multicolumn{1}{c}{$N_3$} & $N_4$ 
\\ \cmidrule{1-7}
$t = 1$   & \multicolumn{1}{c}{$4.6$}    & \multicolumn{1}{c}{$0.5$}     & \multicolumn{1}{c|}{$4.0$}     & \multicolumn{1}{c}{$4$} & \multicolumn{1}{c}{$0$} & \multicolumn{1}{c}{$4$}          
\\ 
$t = 2$  &     \multicolumn{1}{c}{$5.2$}    & \multicolumn{1}{c}{$1.0$}     & \multicolumn{1}{c|}{$6.0$}     & \multicolumn{1}{c}{$5$} & \multicolumn{1}{c}{$1$} & \multicolumn{1}{c}{$6$} 
\\ 
$t = 3$  & \multicolumn{1}{c}{$4.8$}    & \multicolumn{1}{c}{$0.5$}     & \multicolumn{1}{c|}{$4.0$}     & \multicolumn{1}{c}{$4$} & \multicolumn{1}{c}{$0$} & \multicolumn{1}{c}{$4$} 
\\ 
$t = 4$  & \multicolumn{1}{c}{$5.4$}    & \multicolumn{1}{c}{$1.0$}     & \multicolumn{1}{c|}{$6.0$}    & \multicolumn{1}{c}{$5$} & \multicolumn{1}{c}{$1$} & \multicolumn{1}{c}{$6$} 
\\ 
$t = 5$ & \multicolumn{1}{c}{$5.0$}    & \multicolumn{1}{c}{$0.5$}     & \multicolumn{1}{c|}{$5.0$}    & \multicolumn{1}{c}{$5$} & \multicolumn{1}{c}{$0$} & \multicolumn{1}{c}{$5$} 
\\ 
$t = 6$ & \multicolumn{1}{c}{$4.6$}    & \multicolumn{1}{c}{$1.0$}     & \multicolumn{1}{c|}{$4.0$}    & \multicolumn{1}{c}{$4$} & \multicolumn{1}{c}{$1$} & \multicolumn{1}{c}{$5$} 
\\ 

\bottomrule
\end{tabular}
\vspace{10pt}
  \caption{Execution of SNN}    
      \label{tab1}
\end{table}

Table~\ref{tab1} shows the stored membrane potentials for all 6 timesteps, along with the spike amplitudes of each neuron spiking at that time step. The output of  $N_4$ is the tuple $\langle 4, 6, 4, 6, 5, 5\rangle$. According to the definition, the outputs of the SNN are the time average of the cumulative sums i.e. $\langle 4/1, (4+6)/2, (4+6+4)/3, \dots \rangle$ which simplifies to $\langle 4, 5, 4.67, 5, 5, 5 \rangle$.
$\Box$

\subsection{ANN-SNN Conversion}
\noindent
ANN to SNN conversion has been a well-studied problem in recent days~\cite{snnconvgao, snnconvwang, snnconvsengupta}. The conversion algorithms involve input encoding, replacing the neurons of the ANNs with spiking neurons and determining the hyperparameters including the temporal window (NUMSTEPS). Some statistical criterion ensures that the output of the SNN closely mirrors that of the ANN. Typically, the accuracy is in terms of Mean Squared Error (MSE) with respect to a selected dataset.
Since the I/O behaviour and latency of the SNN depend upon the NUMSTEPS hyperparameter, the attempt is to have a translation that achieves the stated accuracy objective with the minimum NUMSTEPS. In this paper, we assume that we already have an SNN with SRLA neurons translated from a trained and verified ANN using the Nengo framework~\cite{Bekolay2014}. Our task is to determine a suitable value for the NUMSTEPS hyperparameter for the SNN.

%% file: sections/formal_model.tex
\section{System Model and Problem Definition}
\label{sec:formal_model}
\noindent
  
\noindent
A range is specified as a real closed interval $[l, u]$, where $l$ is the lower bound and $u$ is the upper bound of the range.
For two ranges $I = [l, u]$ and $J = [l', u']$ denote by $I \le J$ ($I$ is contained in $J$) if $l' \le l$ and $u \le u'$.

Given an ANN $\mathcal{A}$, we denote by $|N_{op}|$ the number of outputs and by $R_i(\mathcal{A})$ the range for the $i$-th output. Let the SNN $\mathcal{N}$ be obtained by an ANN-SNN conversion procedure $Trans(.)$. The SNN has the same number and order of outputs as the ANN $\mathcal{A}$. We denote by $S_i(\mathcal{N})$ the ranges of the $i$-th output of \snn. $MSE_i(D, \ann, \snn)$ denotes the MSE value for the $i$-th output computed using an input dataset $D$ and obtaining the output value by simulating \ann~and \snn. Recall that we denote by $\mathcal{N}_T$ an SNN $\mathcal{N}$ where NUMSTEPS has been set to $T$.

\begin{definition}\textbf{(ANN-SNN Conversion with Safe Range Requirement).}
Given an ANN $\mathcal{A}$, the SNN $\mathcal{N} = Trans(\mathcal{A})$, an input dataset $D$, an MSE bound $\epsilon$, and an integer upper bound $T_{up}$. The problem is to find the smallest timestep $T \le T_{up}$ such that for each $i \in \{1, \dots, |N_{op}|\}$, $MSE_i(D, \mathcal{A}, \mathcal{N}_T) \le \epsilon$ and for each $i$, $S_i(\mathcal{N}_T) \le R_i(\mathcal{A})$.
\hfill $\blacksquare$ 
\end{definition}
\noindent
Note, the smallest timestep is the NUMSTEPS value that we look for. According to the problem definition, the desired SNN controller should have good accuracy (in terms of MSE) relative to the original ANN, and also ensure that the outputs never exceed the bounds of the ANN for any input. Since the ANN is assumed to be safe, 
this ensures that the SNN also remains safe. Note that the accuracy alone is not sufficient for the safe range satisfaction as illustrated next. 

\textbf{Motivational Example:} We take the example of an SNN controller for a double pendulum~\cite{ARCH19} system having two output neurons and MSE objective of 0.4. Table~\ref{tab:motivation} shows the MSE values recorded for the SNN controller for these two output neurons along with the verification results of its safe range requirement. The safe ranges of these two output neurons are 
$R_1(\mathcal{A})=[-5.86571, -3.69253]$ and $R_2(\mathcal{A})=[-6.35836, -3.41698]$.
It can be expected that a lower MSE corresponds to a better convergence of the SNN output towards the ANN output. However, satisfaction of the MSE bound does not ensure the safe range requirement. 
\begin{table}[h]
  \centering
  \renewcommand{\arraystretch}{1.1}
\begin{tabular}{@{}c|cccc@{}}
\toprule
 \textbf{\# of}  & \textbf{MSE} & \textbf{MSE}  & \textbf{Verification}\\ 
\textbf{Timesteps} & \textbf{Output 1} &\textbf{Output 2} & \textbf{Result}\\ 
 \midrule
 $1$ & $4.27391$ & $1.71482$ &  Unsafe\\
 $2$ & $0.54261$ & $0.13999$  & Unsafe\\ 
 $3$ & $0.36017$ & $0.07155$  & Unsafe\\ 
 $4$ & $0.20660$ & $0.04399$ & Safe\\ 
 \bottomrule
\end{tabular}
 \caption{Safety Verification of Double Pendulum}    
      \label{tab:motivation}
\end{table}

\noindent
The SNN controller produces outputs for timesteps $1$ and $2$ with a higher MSE (e.g., $\ge 0.4$), and therefore, can be expected to violate its safe range. However, for the third timestep, the SNN controller produces much lesser MSE values,  
i.e., $0.36$ and $0.07$, corresponding to its two output neurons respectively and satisfies the MSE bound. But still, it fails to meet the given safe range specification. Thus, the possibility of the SNN violating its safe bounds cannot be judged by only data-driven approaches (like evaluating based on MSE) alone. In the example, we find that only in step 4, both the MSE objectives and safe range requirements are met. This elucidates the need for a formal approach to safety verification.

%% file: sections/solutiondetails.tex
\section{Detailed methodology}
\label{sec:detailed_methodology}
\noindent
In this section, we provide a solution for the problem of ANN-SNN conversion with safe range guarantees. First, we give an outline of the overall framework in Algorithm~\ref{alg:main} and then explicate the steps gradually.

\begin{algorithm}
\caption{Find NUMSTEPS for Safe Range Gurantee}
\label{alg:main}
\SetKwInOut{Input}{Input}\SetKwInOut{Output}{Output}
\Input{ANN \ann, 
       SNN \snn, 
       Minimum MSE $\epsilon$, 
       Samples $D$ from the input space, Timestep upper bound $T_{up}$, Safe Ranges $\{[l_i, u_i]\}_{i=1}^{|N_{op}|}$, 
                    for output neurons in $N_{op}$}
\Output{NUMSTEPS, i.e., the smallest timesteps $T$ for the implementation of \snn}
\DontPrintSemicolon
\SetInd{0.3em}{0.7em}

\SetKwFunction{bounds}{SNN\_BOUNDS}
\SetKwFunction{verify}{VERIFY}
\SetKwFunction{mse}{MSE}
{$range \gets \{[l_i, u_i]\}_{i=1}^{|N_{op}|}$} \;
    
\For{$T \gets 1$ to $T_{up}$}{
\If{{\mse{$D$, $\ann$, $\snn_T$} $< \epsilon$}} 
{
\If{\verify{~$\snn_T$, $D$, $range$} is {True}} {\KwRet{$T$}}
\Else{
    {$res \gets$ counterexample from  \texttt{VERIFY()}} \\
    {$snn\_range \gets$ \bounds{~$\snn_T$, $res$}}

    \If{$snn\_range$ is $acceptable$} {\KwRet{$T$}} 
}
}
{\bf display $``$ $No$ $such$ $T$ $is$ $found$ $"$}
}
\end{algorithm}

\emph{Description}: The algorithm takes as input the ANN $\ann$, SNN $\snn$ generated from the ANN, an accuracy objective $\epsilon$, a set $D$ sampled from the input space, and an integer upper bound $T_{up}$ of the timesteps, and lastly a safe range specification.
$T_{up}$ is dependent upon the hardware on which the SNN controller runs and is to be set to $\lfloor{p/e}\rfloor$ where $p$ is the control period of the respective ANN controller and $e$ is the execution time of the SNN \snn~ for a single timestep.

Recall that $N_{op}$ is the set of output neurons of the SNN. The algorithm first sets the safe range of the output neurons of \snn ~(line 1). 
It then iteratively searches for a NUMSTEPS value starting from 1, where the MSE values calculated for all the outputs using the dataset $D$ are less than the accuracy objective $\epsilon$ (line 3). Only then the safe range verification procedure is called (line 4). In the initial iterations, $\snn_T$ is expected to have lesser accuracy, and the possibility of safe range violation is greater. Therefore, line 3 avoids invoking the verification procedure till the MSE converges to $\epsilon$.
If the verification succeeds (i.e., no violation), we return the iteration step as the value for NUMSTEPS (line 5). 
When the verification fails for the safe range requirements derived from the ANN \ann, the verification procedure returns a counterexample $res$, and we opt to compute the actual range supported by $\snn_T$ using $res$ (lines 7-8). Note that a formal verification tool typically reports one counterexample at a time. Since these actual ranges may be larger than the safe ranges derived from the ANN \ann, the CPS developer can validate the ranges of SNN $\snn_T$\ to obtain the required NUMSTEPS (lines 9-10). 

\begin{figure}
     \centering
        \includegraphics[width=1\linewidth]{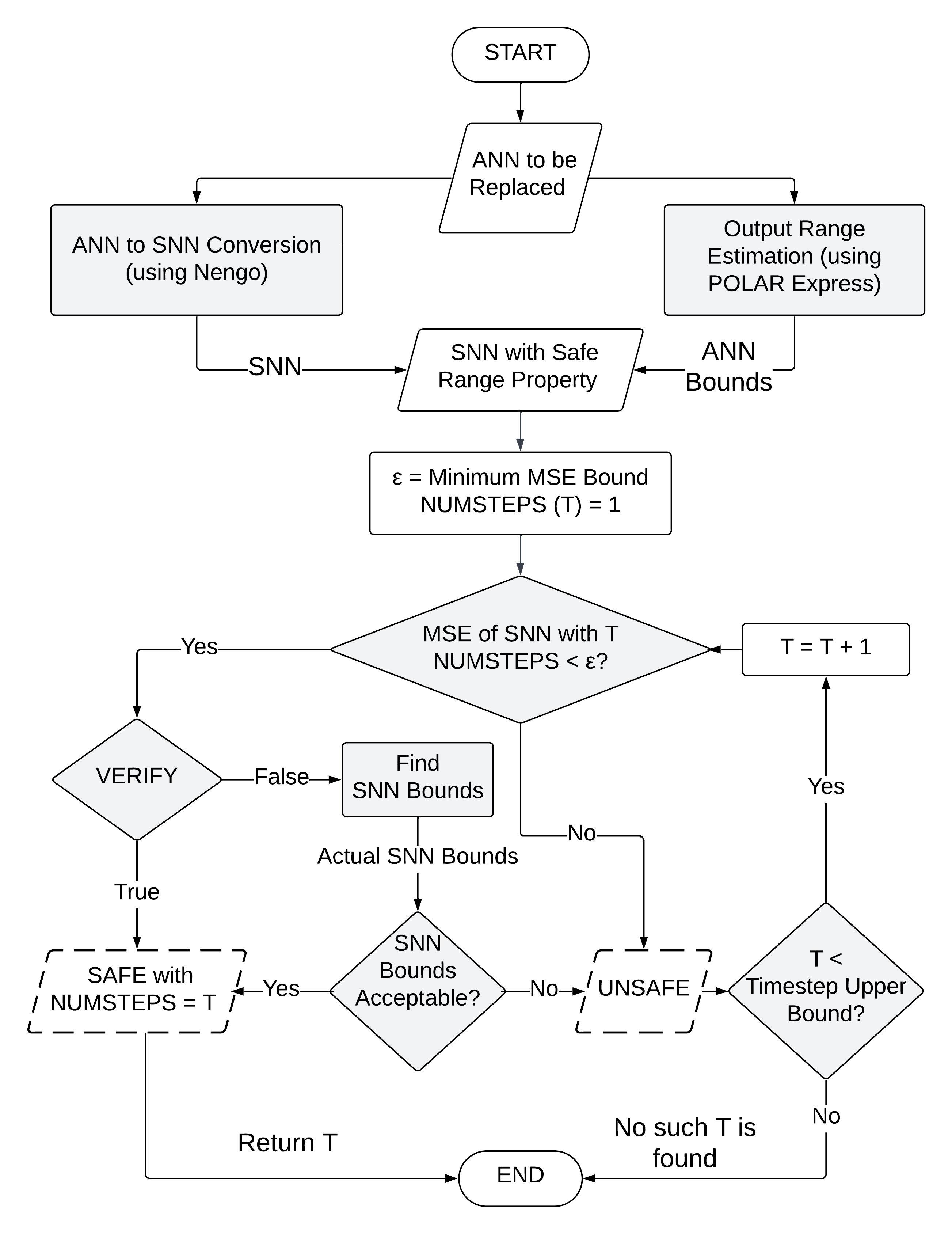}
     \caption{Flow Diagram of the Proposed Methodology}
     \label{fig:fowchart}
 \end{figure}

The safe range specifications in the algorithm are calculated from the original ANN \ann.  As mentioned in Section~\ref{sec:preliminaries}, this can be done using state-of-the-art methods. In this work, we have used the most efficient state-of-the-art reachability analyzer tool POLAR-Express~\cite{polar} to obtain the same. The MSE computation is done from the outputs resulting from simulating the ANN \ann~ and SNN \snn~ on the sampled input $D$. 
Figure \ref{fig:fowchart} depicts the overall flow of the proposed methodology.
In the rest of the section, we give details of the main verification procedure \texttt{VERIFY} and the procedure for computing the actual bounds of the SNN (\texttt{SNN\_BOUNDS}).

\subsection{Verifying the Safe Range Specification for SNN}
\noindent
The verification of the safe range is done only when the MSE requirement is satisfied. Algorithm~\ref{alg:verify} explains the working of the \texttt{VERIFY()} function next.

\begin{algorithm}
\caption{VERIFY} \label{alg:verify}
\SetKwInOut{Input}{Input}
\SetKwInOut{Output}{Output}
\SetInd{0.3em}{0.7em}

\Input{SNN $\mathcal{N_T}$, Input Samples $D$, Safe Range $range$}
\Output{$False$ with a counterexample input in $res$ that violates the safe range spec., $True$ otherwise}
\DontPrintSemicolon
\SetKwFunction{Verify}{Verify}
\SetKwProg{Fn}{procedure}{}{}
\SetKwFunction{simulate}{SIMULATE}
\SetKwFunction{fv}{FV}

\Fn{\verify{$\mathcal{N_T}$, $D$, $range$}}{
\If{\simulate {$\snn_T$, $D$, $range$} returns $no$ $violation$}{
    \If{\fv{~$\snn_T$, $range$} is $True$}{\KwRet{$True$}}
    
}

\KwRet{$False$, $res$}
}

\end{algorithm}

\noindent
\texttt{VERIFY()} consists of two separate steps:
\begin{enumerate}
    \item Simulation: 
    The SNN is simulated using a random subset $D$, chosen from the input space
    to check if there is a safe range violation by some input (returned in $res$ as a counterexample). This is an inexpensive step since it involves executing the given SNN on a specific input, as opposed to formal verification, which requires an exhaustive analysis over all possible inputs.
    If the simulation does not report any violation, only then the more rigorous step of formal verification is invoked.
    
    \item Formal Verification: This involves encoding the SNN in MILP and using an MILP solver to verify if the given safe range for the SNN is satisfied always. When the \texttt{FV()} function returns $True$, the safe range specification for the SNN is ensured for all possible inputs, or $False$ otherwise with a counterexample in $res$. The detail of \texttt{FV()} is given in the next subsection.
\end{enumerate}

%---------------------------------------------------------------------
\subsection{MILP Encoding and Safe Range Verification}
\noindent
The encoding of an SNN starts with the encoding of the execution semantics of the hidden neurons with SRLA on the input coming from the input neurons. 
For all input neurons $N_i$, let $A_{i,t}$ be the real variable denoting the value of the input corresponding to $N_i$ at time $t$. For all other neurons $N_i$, the variable $A_{i, t}$ denotes the integer output of $N_i$ such that $A_{i,t} \geq 0$. 
Note $A_{i,t}$ is the amplitude with which the neuron spikes at time step $t$. 
Let $S_{i,t}$ and $P_{i,t}$ be the real variables denoting the instant potential gained by $N_i$ and the membrane potential of $N_i$ at time $t$. 

\subsubsection{Encoding the Execution Semantics} 
\label{sec:lra_cons}
\noindent
Given an SNN, the execution semantics of its SRLA-based neurons can be formulated using Linear Real Arithmetic (LRA) constraints as follows. 
Note that  $w_{i,j} \in \mathbb{R}$ represents the synaptic weight connecting the adjacent neurons $(N_i, N_j)$, $\theta_{i}$ is the threshold 
of the neuron $N_i$, and $\lambda \in \mathbb{R}$ denotes the \textcolor{black}{leak factor} 
of all the neurons of that SNN.

\par \noindent \textbf{$\mathbf{C_0}$\textit{:~ Initialization of membrane potentials:}} The membrane potential $P_{i,0}$ for each neuron is 
initialized to $0$ at timestep $0$.
\[ C_0(i,0) \triangleq (P_{i,0} = 0) \]

\par \noindent \textbf{\textit{$\mathbf{C_1}$:~ 
The instantaneous potential of a neuron is calculated based on the weights of the incoming synapses (i.e., inSynapse) 
and the amplitudes at which the connected neurons have spiked}:}
During each timestep, when neuron $N_{i}$ receives a spike from neuron $N_{j}$, the product of the synaptic weight $w_{j,i}$ (connecting $N_{j}$ to $N_{i}$) and the spike amplitude $A_{j,t}$ is scaled down by the threshold $\theta_{i}$ of neuron $N_i$ to obtain the incoming potential from neuron $N_j$. The instant potential of neuron $N_i$ is the accumulation of all such incoming potentials at timestep $t$ through each of its incoming synapses.

 \[C_1(i,t) \triangleq (S_{i,t} = \sum_{j \in inSynapse(N_i)} (A_{j,t}\cdot w_{j,i}) /  \theta_{i}) \]
The integer variable $A_{j,t}$, defined such that $A_{j,t} \geq 0$, captures the amplitude of the spike emitted from $N_j$ at time step $t$ and consequently influences the instant potential $S_{i,t}$ of $N_i$, along with the incoming synaptic weight $w_{j,i}$. 

~\\~
\par \noindent \textbf{\textit{$\mathbf{C_2 - C_8}$:~ Encoding the SRLA semantics}:} 
At every time step, an SRLA activation first performs a ReLU operation on the sum of the previously stored membrane potential 
 $P_{i,t-1}$ (after leakage) and the instant potential $S_{i,t}$ (i.e., $S_{i,t} + \lambda \cdot P_{i,t-1}$). 
 The floor operation is then performed on the result of the ReLU operation. The output of the floor function gives the amplitude of the output spike of the neuron. 
Formally, we can encode the SRLA execution semantics as given below. ReLU is a piecewise linear function defined as:
\begin{align*}
% \label{eq:relu}
        ReLU(x) = \left\{\begin{array}{cc}
            x & \ \ \ ;~x>0 \\
            0 & \ \ \ ;~x\leq0  
        \end{array}\right.
\end{align*} 
Now, the SRLA value $A_{i,t}$ for a neuron $N_i$ at timestep $t$ can be given as:
  \begin{align}
    A_{i,t} = \lfloor ReLU(\lambda \cdot P_{i,t-1} + S_{i,t}) \rfloor
\end{align}

\noindent
MILP solvers, to the best of our knowledge, do not have direct support for encoding conditional statements. Hence, we use the big-M method~\cite{Grossmann2002,cheng2017maximum} 
to implement the condition in the ReLU operation.
In the constraints below, the value of the ReLU operation 
is encoded by $x_{i,t}$. 
\begin{equation*}
    \begin{split}
C_2(i, t) & \triangleq  (\left(S_{i,t} + \lambda\cdot P_{i,t-1} + M \cdot q_{i,t} \right) \geq x_{i,t})\\
C_3(i, t) & \triangleq (\left(S_{i,t} + \lambda\cdot P_{i,t-1}\right) \leq x_{i,t}) \\
C_4(i, t) & \triangleq ( x_{i,t} \geq 0) \\
C_5(i, t) & \triangleq ( M \cdot (1 - q_{i,t}) \geq  x_{i,t})\\
C_6(i, t) & \triangleq (0\leq q_{i,t} \leq 1)      
    \end{split}
\end{equation*}
\noindent
Recall that $\lambda$ is the leak factor, as defined in the previous section.  $q_{i,t}$ is the binary variable used for encoding the big-M method.
%$q_{i,t}$ takes values in \{0,1\}. 
$M$ is a sufficiently large number (taken as $+\infty$ in our implementation). 
Next, the constraints $C_7$ and $C_8$ encode the floor operation (not directly supported by MILP solvers) on the result of the ReLU operation, i.e., on $x_{i,t}$. Note, 
$\varepsilon$ is a very small value taken as $-\infty$ in our case for encoding the floor function as follows.
\begin{equation*}
    \begin{split}
C_7(i,t) & \triangleq (A_{i,t} \leq x_{i,t})\\
C_8(i,t) & \triangleq (A_{i,t} + 1 \geq x_{i,t} 
+ \varepsilon)
\end{split}
\end{equation*}

\par 
\noindent \textbf{\textit{$\mathbf{C_9}$:~ The membrane potential of a neuron at any timestep is the amplitude of the emitted spike subtracted from the sum of the instant potential and the previously existing membrane potential scaled by the leak factor}:} 
This is achieved by updating the membrane potential $P_{i,t}$ of neuron $N_i$ at the $t$-th timestep as given next. 
\[C_9(i,t) \triangleq \left(P_{i,t} = (\lambda\cdot P_{i,t-1} + S_{i,t}) - A_{i,t}\right)\]

\par \noindent \textbf{\textit{$\mathbf{C_{10}}$:~ The output of the SNN is the average of a linear sum of the instant potentials at the output neuron(s) over the timesteps}:} 
Let $op_{i}$ be the $i$-th output neuron of the SNN. The value of $op_{i}$ is formulated as follows.
\[ C_{10}(i,T) \triangleq (op_{i} = ( \sum^T_{t=1} S_{i,t} \ / \ T ))\]
Here, $T$ denotes the NUMSTEPS 
over which the average output of the SNN is calculated. 
$S_{i,t}$ is the instant potential gained at the $i$-th output neuron at timestep $t$. 
The MILP encoding $F_{\mathcal{N}_{T}}$ of a given SNN is the conjunction of $C_0, C_1,C_2,\ldots C_{10}$ as stated below, across all neurons over NUMSTEPS $T$.

\begin{equation}
\label{eq:snn}
\begin{aligned}
& F_{\snn_T} \triangleq ~ 
  \biggl(\bigwedge_{i \in N} C_0(i, 0)\biggl)  \\
  & \mand \biggl(\bigwedge_{t\in \{1\ldots T \}} \bigwedge_{i \in N} C_1(i, t)   \mand C_2(i, t) \\ & \mand C_3(i,t) 
  \mand C_4(i,t) \mand C_5(i,t) \mand C_6(i,t)  \\ & \mand C_7(i,t) \mand C_8(i,t) \mand C_9(i,t) \biggl) \\ &
  \mand \biggl(
  \bigwedge_{i \in N_{op}} C_{10}(i, T)\biggl)
\end{aligned}
\end{equation} 
\noindent
Here, $N$ is the set of all neurons with SRLA and $N_{op}$ is the set of all output neurons in the SNN.

\subsubsection{Encoding the Safe Range Spec.}
\noindent
Given the MILP encoding of the SNN, our next task is to encode the {\em negation of the safe range requirement} so that we can verify its satisfaction. Recall that we are given the safe ranges for all the output neurons in the SNN, i.e, $\{[l_i, u_i]\}_{i=1}^{|N_{op}|}$,
where $|N_{op}|$ is the total number of output neurons. 
We check for satisfaction against these two bounds with two separate queries. 
The query for checking the negation of the safe upper bound is encoded as: 
\begin{equation}\label{qub}
    \psi_{ub} \triangleq F_{\snn_T} \mand (op_{i} \geq u_i ).
\end{equation}
Note that $op_{i}$ is the output value of the $i$-th output neuron.
Similarly, we formulate the verification query for checking the negation of the safe lower bound of the $i$-th neuron. % $Q_{lb}$.  
\begin{equation} \label{qlb}
    \psi_{lb} \triangleq F_{\snn_T} \mand (op_{i} \leq l_i )
\end{equation}
With both queries, we search for an input that violates the given safe range specification. It holds only if both the above queries are proven infeasible by the solver, indicating that no input can trigger an output that is outside the safe range.
We can extend Eq.\eqref{qub} (and similarly Eq.~\eqref{qlb}) for checking the upper bounds of all the output neurons in $N_{op}$ as follows.
\begin{equation}\label{qub2or}
\begin{aligned}
\psi_{ub} \triangleq  F_{\snn_T}  \mand
  \biggl(\bigvee_{i \in N_{op}} (op_{i} \geq u_i)  \biggl)
\end{aligned}
\end{equation} 

\noindent 
It is important to note that, since the disjunction operator is not supported by MILP solvers, we use the big-M method to encode this disjunctive condition as follows. 
\begin{equation}\label{qub2}
\begin{aligned}
&\psi_{ub} \triangleq  F_{\snn_T}  \mand  
  \biggl(\bigwedge_{i \in N_{op}} (op_{i} - u_i \leq  M\! \cdot\! x_i) \mand \\   
 & (u_i - op_{i} \leq  M\!\! \cdot\! (1 - x_i))   
  \!\!\mand\!\! (x_i \in \{0,1\}) \biggl)
  \!\!\mand\!\! \\ 
&  \biggl(\sum^{|N_{op}|}_{i = 1} x_i \geq 1 \biggl) 
\end{aligned}
\end{equation} 
\noindent Each variable $x_i$ is satisfied with a value of 1 when an input violating the upper bound is found. 
We can use a similar query to verify the lower bounds of all the output neurons in $N_{op}$. %of the SNN.

\subsubsection{Formal Verification} 

\begin{algorithm}
\caption{FV}
\label{alg:FV}

\SetKwInOut{Input}{Input}
\SetKwInOut{Output}{Output}
\SetInd{0.3em}{0.7em}

\Input{The SNN that runs for T timesteps, $\snn_T$, Safe range for SNN outputs, i.e., $range=\{[l_i, u_i]\}_{i=1}^{|N_{op}|}$}     
\Output{$False$ with a counterexample input in $res$ that violates the safe range, $True$ otherwise}

\DontPrintSemicolon
\SetKwProg{Fn}{procedure}{}{}
\SetKwFunction{fvub}{FV\_UB}
\SetKwFunction{fvlb}{FV\_LB}
\SetKwFunction{fv}{FV}
\SetKwFunction{enc}{ENCODE}

\Fn{\fv{$\snn_T, range$}}{
$F_{\snn_T} \gets$ \enc{$\snn_T$} 

$L \gets \{l_i\}_{i=1}^{|N_{op}|}$, \ \ $U \gets\{u_i\}_{i=1}^{|N_{op}|}$ \;
\If{\fvlb{$F_{\snn_T}$, $L$} returns $False$}{
 \KwRet{$False$, $res$}
}
\If{\fvub {$F_{\snn_T}$, $U$} returns $False$}{
 \KwRet{$False$, $res$}
}

\KwRet{$True$}
%\EndProcedure

}
\end{algorithm}

\noindent
Finally, \texttt{FV()} outlined in Algorithm~\ref{alg:FV} first encodes the SNN $\snn_T$ following the formulation given in Eq.~\eqref{eq:snn} (line 2). Next, it calculates the lower bound $L$ and the upper bound $U$ from the safe range, $range$, given as input to it. It then invokes the \texttt{FV\_LB()} and \texttt{FV\_UB()} (details given next) for verifying the safe output bounds (lines 4,6). When any of these functions returns $False$, indicating the violation of the safe range, \texttt{FV()} terminates and returns $False$ with the counterexample, otherwise, it returns $True$. 

\begin{algorithm}
\caption{FV\_UB}\label{alg:FV_UB}

\SetKwInOut{Input}{Input}
\SetKwInOut{Output}{Output}
\SetInd{0.3em}{0.7em}

\Input{The encodings of the SNN which runs for T timesteps i.e., $F_{\snn_T}$, Safe upper bounds, $U=\{u_i\}_{i=1}^{|N_{op}|}$
        }
        
\Output{$False$ when an input that violates the safe upper bound exists, $True$ otherwise}

\SetKwProg{Fn}{procedure}{}{}
\SetKwFunction{qub}{$Q_{ub}$}
\SetKwFunction{fvub}{FV\_UB}
\DontPrintSemicolon

\Fn{\fvub{$F_{\snn_T}$, $U$}}{

$\psi^T_{ub} \gets$ \qub{$F_{\snn_T}$, $U$} \;%following Eq.\eqref{qub2} 
\If{$\psi^T_{ub}$ is $feasible$}{
\KwRet{$False$}}
\Else {
\KwRet{$True$}
} 
}
\end{algorithm}

\texttt{FV\_UB()} in Algorithm~\ref{alg:FV_UB} uses the \texttt{$Q_{ub}()$} function internally to encode the upper output bound for the safety check using the formulation given in Eq.~\eqref{qub2} or Eq.~\eqref{qub} (required when we run \texttt{FV\_UB()} for one neuron at a time, as used in Algorithm~\ref{alg:bound_tightening}). Similarly \texttt{FV\_LB()} is used for the lower bound check.

\subsection{Binary Search for Iterative Bound Tightening}
\label{sec:iterative_bound_tightening}
\noindent
When we find that the given safe range specification of the SNN is violated, we compute the actual range supported by the SNN outputs employing the function \texttt{SNN\_BOUNDS()}. However, in our experiments, we find that computing the SNN output range through an objective function often leads to timeouts. Hence, we propose a binary search algorithm that uses verification procedures \texttt{FV\_UB()} and \texttt{FV\_LB()} as subroutines, to estimate tight ranges within a factor $\delta$ of the actual ranges obtained during the safe range verification in Algorithm~\ref{alg:verify}.

\texttt{SNN\_BOUNDS()} is implemented using two functions \texttt{FIND\_UB()} and \texttt{FIND\_LB()} to tighten the upper and lower bounds of the outputs of a given SNN $\snn_T$ respectively. 
Algorithm~\ref{alg:bound_tightening} outlines \texttt{FIND\_UB()}.  
The algorithm for \texttt{FIND\_LB()} is similar in nature.  
Recall that in line 7 of Algorithm~\ref{alg:main}, we store the counterexample in $res$, returned by the \texttt{VERIFY()} procedure in Algorithm~\ref{alg:verify}. The value stored in $res$ is the input for which the safe range gets violated. \texttt{FIND\_UB()} starts by setting the variable $U_{i}^{ce}$ of the $i$-th output neuron, to the actual upper bound it gets by executing the counterexample in $res$. 
The variable $U_{i}^{vio}$, of the $i$-th output neuron, is initialized with the value of the variable $U_{i}^{ce}$ and then updated to the highest observed output value of the $i$-th output neuron (found from the last encountered counterexample).
The value of $U_{i}^{ce}$ is then incremented iteratively by a value of $\beta$ and then verified using the procedure \texttt{FV\_UB()} for $K$ iterations till it returns $True$ (here, we run \texttt{FV\_UB()} with Eq.~\eqref{qub} only).
This ensures an upper bound supported by the SNN $\snn_T$.
The value for $\beta$ was chosen after multiple trials.
At this point, the least upper bound lies in the interval $(U_{i}^{vio} (left)$, $U_{i}^{ce} (right)]$ (lines 9-10). The interval is tightened by using a binary search between these two values.

\begin{algorithm}
\caption{FIND\_UB}
\label{alg:bound_tightening}
\SetKwInOut{Input}{Input}
\SetKwInOut{Output}{Output}
\SetInd{0.3em}{0.7em}

\Input{Encoding $F_{\snn_T}$ of SNN $\snn_T$, %List of $\mathcal{L}$
Value of $T$ for which MSE $<$ $\epsilon$, Safe upper bound of $i$-th output neuron $u_{i}$, Iteration bound $K$, parameters $\beta$ and $\delta$ 
}
        
\Output{The tightened upper bound $U_{i}^{tgt}$  %, tightness of the upper bound achieved $tightness$
}

\SetKwProg{Fn}{procedure}{}{}
\SetKwFunction{qub}{$Q_{ub}$}
\SetKwFunction{findub}{FIND\_UB}
\SetKwFunction{fvub}{FV\_UB}
\SetKwFunction{simulate}{SIMULATE}
\SetKwFunction{rand}{RAND}

\DontPrintSemicolon

\Fn{\findub{$F_{\snn_T}$, $\mathcal{L}$, $U_{i}$}}{ 
\tcp{Incrementing the bound until \texttt{FV\_UB()} returns $True$}

$U_{i}^{ce} \gets u_{i}$

\For{$k \gets 1$ to $K$}{
    $U_{i}^{vio} \gets U_{i}^{ce}$ \\
    $ U_{i}^{ce} \gets U_{i}^{ce} + \beta$ \; 
    %\State $ U_{i}^{ce} \gets U_{i}^{ce} + \beta$
    $ ans \gets$ \fvub{$F_{\snn_T}, U_{i}^{ce}$}

\If{$ans$ is $True$}{
$break$
}
}

\tcp{Binary Search to find a tightened bound}
$left \gets U_{i}^{vio}$ \\
$right \gets U_{i}^{ce}$ \;
\While{$(right - left) > \delta $}{
$mid \gets$ \rand{$left+\delta, right$} \;
\If{\simulate{$\snn_T, D$, $mid$} returns $a$ $violation$}{
    $left \gets mid$
    }
\ElseIf{\fvub{$F_{\snn_T}, mid$} is $False$ \textbf{or} $Timeout$}{ 
    $left \gets mid$
}
\Else {
    $right \gets mid$
}

}

$U_{i}^{tgt} \gets right$ \;

\KwRet{$U_{i}^{tgt}$}

}
\end{algorithm}

We use a variant of the binary search algorithm, namely, {\em random binary search}~\cite{randbs}.
The function \texttt{RAND(.)} on line 12 generates a random real value $mid$ within the range $(left+\delta, right)$. 
Lines 13-16 check if the value of $mid$ is not an appropriate upper bound of the $i$-th output neuron of $\snn_T$ through a random simulation and a formal \texttt{FV\_UB($F_{\snn_T}, mid$)}, in which case the $left$ bound of the interval is updated to $mid$. When \texttt{FV\_UB($F_{\snn_T}, mid$)} has a Timeout, we conservatively increment $left$ to $mid$. On the other hand, if \texttt{FV\_UB($F_{\snn_T}, mid$)} is True (lines 17-18), it implies we have got a better upper bound than the current $right$, hence it is updated to $mid$. The while loop (11-18) stops when we get an upper bound ($right$) which is within $\delta$ of the lower bound ($left$) and returns $right$ as the $\delta$-tight upper bound $U_{i}^{tgt}$. The value of $\delta$ is user-specified and we use the value of $0.001$ for our experiments.

%% file: sections/experiment.tex
\section{Implementation and Results} 
\label{sec:implementation}
\noindent
This section describes the implementation details followed by the experimental results.  
\begin{table*}[t]
  \centering
  \renewcommand{\arraystretch}{1.1}
\begin{tabular}{@{}|c|c|c|c|c|c|c|@{}} 
\toprule

\textbf{\makecell{Benchmarks}} & \textbf{\makecell{SNN \\ Architecture}} & \textbf{\makecell{$T_{up}$}}
&\textbf{\makecell{Initial Input \\ Conditions}} & \textbf{\makecell{Safe Range for \\ SNN Controllers}} & \textbf{\makecell{Queries to be\\ Verified}} \\
\midrule

\makecell{Linear Inverted \\ Pendulum} & $4 \times 10 \times 1$ & $25$ & \makecell{$x0 = [-0.5, 0.5], x1 = 0.0,$ \\ $x2 = [-0.2, 0.2], x3 = 0.0$} & $[-15.50883, 15.34465]$ & \makecell{$(op \leq -15.50883) ~ \vee $ \\ $(op \geq  15.34465)$} \\

\midrule

\makecell{Double \\ Pendulum} & $4 \times 25\times 25 \times 2$ & $25$ & \makecell{$x0 = [1.0, 1.3],$ \\$ x1 = [1.0, 1.3],$ \\ $x2 = [1.0, 1.3],$ \\ $x3 = [1.0, 1.3]$} & \makecell{$[-5.86571, -3.69253],$ \\ 
$[-6.35836, -3.41698]$} & 
\makecell{$(op_1 \leq  -5.86571)$ $ \vee $  \\ $(op_2 \leq  -6.35836)~ \vee $ \\  $(op_1 \geq  -3.69253)$ $ \vee $  \\ $(op_2 \geq  -3.41698)$} \\

\midrule

\makecell{Single \\ Pendulum} & $2 \times 25 \times 25 \times 1 $ & $20$ & \makecell{$x0 = [1.0, 1.2], x2 = [-0.2, 0.2]$} & $[-0.78130, -0.54282]$ & \makecell{$(op \leq -0.78130) ~ \vee $ \\ $(op \geq  -0.54282)$} \\

\midrule

ACC3 & $5\times 20\times 20\times 20 \times 1$ & $5$ &\makecell{$x0 = 30, x1 = 1.4,$ \\ $x2 = [90, 110], x3 = [10, 11],$ \\ $x4 = [32, 32.2], x5 = [30, 30.2]$} & $[-1.46030, -0.73179]$ & \makecell{$(op \leq -1.46030) ~ \vee $\\ $(op \geq  -0.73179)$} \\

\midrule

ACC5 & $5 \times 20 \times 20 \times20 \times20 \times20 \times 1 $ & $5$ & \makecell{$x0 = 30, x1 = 1.4,$ \\ $x2 = [90, 110], x3 = [10, 11],$ \\ $x4 = [32, 32.2], x5 = [30, 30.2]$} & $[-8.81953, -6.85014]$ & \makecell{$(op \leq -8.81953)~ \vee $ \\ $(op \geq  -6.85014)$} \\

\bottomrule
\end{tabular}
\caption{Details of Different Benchmarks}
\label{table_polar_prop}
\end{table*}

\paragraph{Benchmarks}
In our work, we consider $5$ benchmark ANN controllers collected from the competition at the ARCH workshop in the years $2019$~\cite{ARCH19} and $2022$~\cite{ARCH22} where the networks were provided by academia as well as industry. These include controllers for a linear inverted pendulum (LIP), a double pendulum (DP), a single pendulum (SP),  and adaptive cruise controllers with $3$ (ACC3) and $5$ hidden layers (ACC5). The architecture of each benchmark is given in Column 2 of Table~\ref{table_polar_prop}. For example, in the case of ACC3, there are $5$ input neurons, $3$ hidden layers having $20$ neurons each, and $1$ output neurons. Note, the double pendulum has two output neurons, while the others have only one output neuron.

\paragraph{Experimental Setup}
All experiments were carried out on Windows 10 running on an Intel Core i7 CPU with 1.30 GHz clock speed and 16 GB of RAM. We used POLAR-Express (commit 13d42b0) downloaded from GitHub on  Aug. 18, 2023, for calculating the safe range of the above-mentioned ANN controllers and the Gurobi MILP solver (version 10.0.3) downloaded from the official website, for constraint solving and formal verification. 

\newcolumntype{x}[1]{>{\centering\arraybackslash\hspace{0pt}}p{#1}}
\newcolumntype{j}[1]{>{\centering\arraybackslash\hspace{0pt}}p{#1}}

\begin{figure}[ht]
  \begin{subfigure}{0.24\textwidth}
  \centering
    \includegraphics[width=\linewidth]{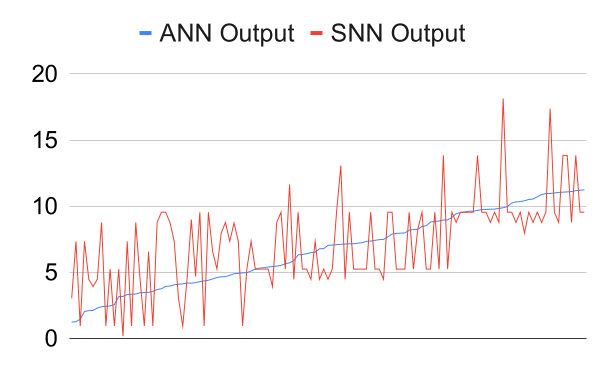}
    \caption{For 1 NUMSTEPS}
    \label{fig:1a}
  \end{subfigure}%
  \hspace*{\fill}   
  \begin{subfigure}{0.24\textwidth}
    \includegraphics[width=\linewidth]{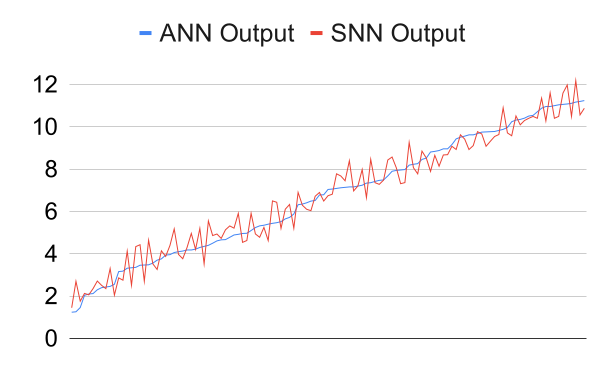}
    \caption{For 5 NUMSTEPS}
    \label{fig:1b}
  \end{subfigure}

  \hspace*{\fill}   
  \par
  \centering
  \begin{subfigure}{0.24\textwidth}
    \includegraphics[width=\linewidth]{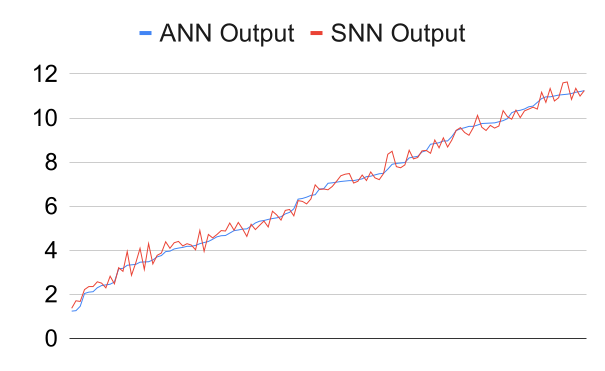}
    \caption{For 10 NUMSTEPS}
    \label{fig:1c}
  \end{subfigure}%
  \hspace*{\fill}   
  \centering
  \begin{subfigure}{0.24\textwidth}
    \includegraphics[width=\linewidth]{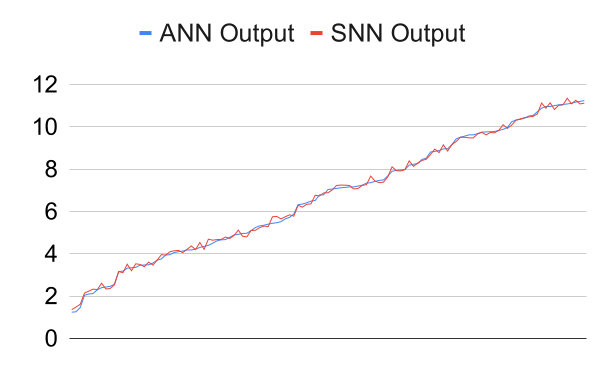}
    \caption{For 25 NUMSTEPS} \label{fig:1d}
  \end{subfigure}
  \hspace*{\fill}   

\caption{\centering Outputs of ANN Controller vs. SNN Controller for Linear Inverted Pendulum with increasing NUMSTEPS} 
\label{fig:LIP_ann_vs_snn}
\end{figure}

\paragraph{ANN-SNN Conversion}
For our experiments, we used the Nengo~\cite{Bekolay2014} framework for translating SNNs from the ANNs.
Nengo accomplishes the transformation by replacing the ReLU activation with SRLA, while preserving the same network architecture, including weights and biases.
For any neuron, the default value of the threshold in Nengo is $1$. 
Figure~\ref{fig:LIP_ann_vs_snn} presents a comparison of outputs produced by an ANN and the corresponding SNN obtained using Nengo, on a randomly sampled set of inputs for the LIP system. It can be observed from Figure~\ref{fig:LIP_ann_vs_snn} that the output of the SNN converges towards the ANN output (for the same input values) as we increase the value of NUMSTEPS. This shows that the MSE gets reduced with increasing value of NUMSTEPS.

\paragraph{Computing Safe Range and Upper bounds on NUMSTEPS}
We calculate the safe range specifications for all five benchmark ANN controllers by running the POLAR-Express tool~\cite{polar} on them. The safe ranges are reported in Column 5 of Table~\ref{table_polar_prop}. It is important to note that these safe ranges are calculated based on some initial conditions of the input variables as reported in Column 4 of the table. 
In particular, we extract the safe bounds of the ANN controllers after a single plant-control iteration and set these as the safe range specifications to be satisfied by the corresponding SNN controllers. Note we can do the same considering all the plant-control iterations towards the satisfaction of a given safety property, integrating plants and SNN controllers. However, this is currently out of the scope of this paper. We aim to do it in the future. 
Now, the queries to be verified by the solver are the negation of these safe ranges as reported in Column 6 of the table. 
For each of these five benchmark systems, the upper bound, $T_{up}$, on NUMSTEPS for SNN controllers are reported in Column 3 of Table~\ref{table_polar_prop}. These values are calculated from the control periods given in~\cite{ARCH19, ARCH22} and by assuming the execution time of the respective SNN for a single timestep as \SI{0.002}{\second} for LIP, \SI{0.008}{\second} for DP, \SI{0.0025}{\second} for SP, and \SI{0.02}{\second} for the ACCs.

%% file: sections/results.tex
\begin{table*}[t]
  \centering
  \scriptsize
  \renewcommand{\arraystretch}{1.1}
\begin{tabular}{@{}|c|c|c|c|c|c|@{}} \hline 

\toprule
\textbf{\makecell{Benchmarks}} & \textbf{\makecell{\# of \\ Timesteps}}  & \textbf{\makecell{MSE}}  & \textbf{\makecell{Verification \\ Result}} &\textbf{\makecell{Range Obtained \\ from Bound Tightening}} & 
\textbf{\makecell{Total Time \\ Taken}}\\ 

\midrule

\multirow{10}{*}{Linear Inverted Pendulum} & $1$ & $13.03031$ & $-$ & $-$ & $-$\\ 

& $2$ & $3.32945$ & $-$ & $-$ & $-$ \\
& $3$ & $1.45138$ & $-$ & $-$ & $-$ \\
& $4$ & $0.81444$ & $-$ & $-$ & $-$ \\
& $5$ & $0.52856$ & $-$ & $-$ & $-$ \\
& $6$ & $0.35786$ & $-$ & $-$ & $-$ \\
& $7$ & $0.26885$ & $-$ & $-$ & $-$ \\
& $8$ & $0.20333$ & $-$ & $-$ & $-$ \\
& $9$ & $0.16450$ & $-$ & $-$ & $-$ \\
& $10$ & $0.13235$  & Safe & $-$ & $0.071s$\\

\midrule

\multirow{6}{*}{Double Pendulum} & $1$ & $4.27391, 1.71482$ & $-$ & $-$ & $-$\\   

& $2$ & $0.54261$, $0.13999$ & $-$ & $-$ & $-$\\   

 & $3$ & $0.36017$, $0.07155$ & $-$ & $-$ & $-$\\ 

 & $4$ & $0.20660, 0.04399$ & $-$ & $-$ & $-$\\ 

 & $5$ & $0.12962, 0.02461$ & $-$ & $-$ & $-$\\ 

 & $6$ & $0.08968, 0.02089$  & Safe & $-$ & $3.736s$\\

\midrule

\multirow{20}{*}{Single Pendulum} & $1$ & $0.43389
$ & $-$ & $-$ & $-$\\   

  & $2$ & $0.40772$ & $-$ & $-$ & $-$\\   

  & $3$ & $0.16405$ & $-$ & $-$ & $-$\\   
  & $4$ & $0.04138$ & Unsafe & $[-0.78130, -0.29494]$ & $0.187s$\\   
  & $5$ & $0.03049$  & Unsafe & $[-0.78130, -0.36013]$ & $0.276s$\\   
  & $6$ & $0.01319$ & Unsafe & $[-0.78130, -0.42488]$ & $0.846s$\\   
  & $7$ & $0.00686$ & Unsafe & $[-0.78130, -0.48474]$ & $1.716s$\\   
  & $8$ & $0.00834$ & Unsafe & $[-0.78130, -0.42463]$ & $2.221s$\\   
  & $9$ & $0.00773$ & Unsafe & $[-0.78130, -0.44675]$ & $15.336s$\\   
  & $10$ & $0.00696$ & Unsafe & $[-0.78130, -0.50055]$ & $68.382s$\\   
  & $11$ & $0.00551$ & Unsafe & $[-0.78130, -0.47896]$ & $90.027s$\\   
  & $12$ & $0.00631$ & Unsafe & $[-0.78130, -0.44960]$ & $55.715s$\\   
  & $13$ & $0.00620$ & Unsafe & $[-0.78130, -0.46947]$ & $112.160s$\\   
  & $14$ & $0.00467$ & Unsafe & $[-0.78130, -0.45901]$ & $304.674s$\\   
  & $15$ & $0.00341$  & Unsafe & $[-0.78130, -0.43867]$ & $319.812s$\\   
  & $16$ & $0.00206$ & Unsafe & $[-0.78130, -0.47342]$ & $1298.872s$\\   
  & $17$ & $0.00141$ & Unsafe & $[-0.78130, -0.49071]$ & $4298.335s$\\   
  & $18$ & $0.00106$ & Unsafe & $[-0.78130, -0.50608]$ & $5850.720s$\\   
  & $19$ & $0.00099$ & Unsafe & $[-0.78130, -0.52872]$ & $5673.547s$\\   
  & $20$ & $0.00103$ & Unsafe & $[-0.78130, -0.51042]$ & $9962.996s$\\   
\midrule
 & $1$ & $0.01412$ & $-$ & $-$ & $-$\\   
 & $2$ & $0.00356$ & $-$ & $-$ & $-$\\   
ACC3 & $3$ & $0.00164$  & Unsafe & $[-1.47700, -0.61709]$ & $424.906s$\\   
 & $4$ & $0.00090$ & Unsafe & $[-1.46733, -0.64829]$ &  $2817.854s$\\   
 & $5$ & $0.00058$ & Unsafe & $[-1.46616, -0.66411]$ & $7902.435s$\\  

\midrule
 & $1$ & $0.00590$ & $-$ & $-$ & $-$\\   
 & $2$ & $0.00145$ & $-$ & $-$ & $-$\\   
ACC5 & $3$ & $0.00070$  & Unsafe & $[-8.82560, -6.81908]$ & $1019.328s$\\   
 & $4$ & $0.00038$ & Unsafe & $[-8.91954, -6.81121]$ & $23918.844s$\\   
 & $5$ & $0.00024$  & Unsafe & $[-8.90126, -6.78736]$ & $15508.534s$\\  
\bottomrule
\end{tabular}
\caption{\centering Safe Range Verification Results of SNN-Controllers}
\label{TableAll}
\end{table*}

\paragraph{Results and Analysis}
We run our framework as described in Algorithm~\ref{alg:main} for each of the five benchmark controllers. 
We consider the minimum MSE bound $\epsilon$ for LIP, DP, SP, ACC3, and ACC5 as $0.15$, $0.10$, $0.05$, $0.002$ and $0.001$ respectively. 
For our experiments, we chose these MSE bounds within 1\% envelope of the safe range specifications.
Table \ref{TableAll} summarizes the overall results. 
Column 3 shows the MSE recorded for each of the SNN controllers based on $5000$ random samples while running our framework.  
As per Algorithm~\ref{alg:main}, we proceed with the verification only when the MSE values obtained by the SNN controllers are below the given minimum bound $\epsilon$. 
Column 4 indicates the verification answer for the given safe range specification. 
Note that we do not verify the safe ranges for timesteps up to $9$, $5$, $3$, $2$ and $2$ for LIP, DP, SP, ACC3, and ACC5 respectively as they do not meet their desired accuracies (in terms of $\epsilon$), hence, the respective cells in the table are marked by --. 
For LIP and DP, the respective SNN controllers satisfy the given safe range requirements. Therefore, for these two systems, we obtain the NUMSTEPS values as $10$ and $6$ respectively, and therefore, we do not verify with higher values, following Algorithm~\ref{alg:main} (lines 4-5). These NUMSTEPS values (i.e., $10$ and $6$) are quite small, and hence, suitable for SNN implementation in practice. 

However, for the other 3 systems, though the respective SNN controllers satisfy the MSE criteria, they fail the safe range verification with all timesteps up to $T_{up}$. This is obtained through only $500$ random input simulations as described in line 2 of the \texttt{VERIFY()} function (i.e., Algorithm~\ref{alg:verify}), before the call to \texttt{FV()} for formal verification in line 3. Thus, we avoid the costly FV calls for SP, ACC3 and ACC5.
Thereafter, for these SNN controllers, an attempt is made to tighten ranges over their actual ranges computed from the counterexample returned by the solver on running Algorithm~\ref{alg:verify}. For all the unsafe instances of these SNN controllers, the tight ranges obtained after running Algorithm~\ref{alg:bound_tightening} (with $K=5$, $\delta=0.001$, and $\beta=0.01$), are shown in Column 5.

\begin{figure}[h]
     \includegraphics[width=0.9613\linewidth]
     {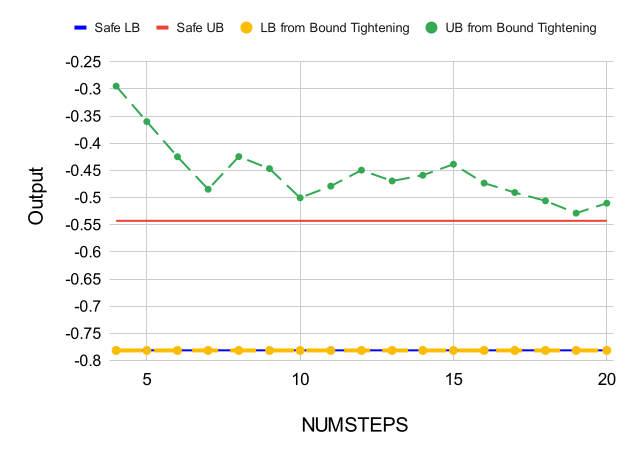}
     \caption{\centering Given Safe Bounds vs. Bounds obtained for SNN Outputs for Single Pendulum across Timesteps}
     \label{fig:graph_SP}
 \end{figure}

 Particularly, for SP, Figure~\ref{fig:graph_SP} compares the given safe ranges (solid red and blue lines for upper and lower bounds respectively) with the tight ranges (dotted green and dotted yellow lines for upper and lower bounds respectively) obtained by running Algorithm~\ref{alg:bound_tightening} for $4$ to $20$ timesteps. 
 
It is worth noticing that the blue and yellow lines coincide indicating the fact that the lower bounds of the SNN outputs for all these timesteps are exactly the same as the given safe lower bounds, which means, they are never violated. Whereas, the distance between the red and green lines indicates the deviations occurred from the given safe upper bounds, leading to a violation of the given safe range. However, this graph shows an overall decreasing trend in the deviation with increasing NUMSTEPS.
These tightened ranges of the SNN output 
will act as feedback to the designer for further improvement of SNN attributes (e.g., stepsize).
Column 6 reports the total time taken to run Algorithm~\ref{alg:main} for each of these five cases. 

%% file: sections/related.tex
\section{Related Work}
\label{sec:related}
\noindent
We categorize the relevant existing works into three main directions as described below.

\par\textit{Safety Verification of ANN-Controlled CPS:}
There exists a rich body of literature that focuses on the safety verification of ANN-controlled CPSs. 
The works proposed in~\cite{reachSherlock, polar} leverage the Taylor model approximation of the ANN controllers 
to compute the over-approximated reachable sets, whereas \cite{2020reachnn} and \cite{hoang_19} do the reachability analysis using Bernstein polynomials and computing star sets respectively. In contrast, \cite{verisig2} converts the ANN controller to an equivalent nonlinear hybrid system combined with the plant dynamics to verify the given safety properties.
Apart from these works, some research efforts specifically focus on the approximating output range of the ANNs~\cite{Pavithra, Katz2017ReluplexAE, dutta2017output} without integrating plants in the loop.  

\par\textit{ANN to SNN Translation:} 
A handful of research works exist that aim to achieve a lossless conversion of ANNs to SNNs. The main idea is to replace the ANN-based controller with an energy-efficient SNN-based system with almost no change in the behaviour. The method proposed in \cite{snnconvgao} is the first approach towards this ANN-SNN conversion and it is particularly applicable for object recognition. The method proposed in~\cite{snnconvwang} introduces a new activation called StepReLU (similar to ReLU) for training an ANN which is more compatible with SNNs using membrane potential encoding. 
A translation method that scales to very deep convolutional as well as residual architectures is given in~\cite{snnconvsengupta}. This method involves layer-by-layer threshold balancing to achieve a near-lossless ANN-SNN conversion. \cite{snnconvdiehl} develops data-based and model-based methods for normalization. Similarly, methods like ~\cite{snnconvrueckauer, snnconvding} aim for near-lossless conversion by parameter (weights and biases) normalization. 

\par\textit{Formal Modeling and Verification of SNNs and SNN-Controlled CPS:}
Although a significant amount of research efforts exists focusing on the formal modeling and verification of ANNs~\cite{Katz2017ReluplexAE,Pavithra}, very few efforts do the same for SNNs~\cite{snnmv2018,vmcai}. The first approach towards the formal modeling of SNNs is found in~\cite{snnmv2016} where an SNN with weighted synapses is transformed into a type of timed safety automata. On the other hand, \cite{snnmv2018} encodes SNNs using timed automata where each neuron is modeled as an individual automaton operating in parallel and interacting through shared channels, depending on the network structure. 
Since both papers utilize techniques from timed automata for modeling and verification, they suffer from the inherent scalability issue of timed automata and, hence not usable in practice specifically in the CPS domain.
A recent attempt to model SNNs with the LIF activations using satisfiability modulo theory is found in~\cite{vmcai}. However, this work specifically focuses on safety verification related to classification problems, whereas most of the neural controllers in practice are for regression problems. Moreover, it fails to scale for the larger SNNs used as controllers as we explored in our experiments.
The temporal nature of the computation in the SNNs bears similarity with the SMT-based approach for verification of NN-based controllers~\cite{AmirSK21}. However, the additional structure of the SNNs, e.g., the spiking outputs and the absence of the environment models has the potential for better scalability.

The safety verification of SNN-controlled CPS is a relatively less explored area of research. %hence, there exist quite a few works. 
The only work in this direction is the approach given in~\cite{partharoop2023}. Here, the authors leverage timed automata for modeling both the plant and the SNN controller and pose the safety verification as the query to be checked by the timed automata. They train SNNs with three types of neurons, namely, LIF, QIF and Izhikevich. The primary limitation is the scalability issue as also admitted by the authors. As an alternative to overcome such issues, they propose to use the statistical verification method. In contrast, in this work, our framework is capable of handling larger SNNs that are compatible with practical CPS benchmarks. Moreover, unlike ~\cite{partharoop2023}, we translate these SNNs from existing safe ANNs, having minimal temporal windows for these SNNs so that a balance between the accuracy and the computational cost can be guaranteed, which is missing in~\cite{partharoop2023}. 

%% file: sections/conclusion.tex
\section{Conclusion and Future Work} \label{sec:conclusion}
\noindent 
This work focuses on the design of safe SNN controllers applicable to safety-critical CPSs. One way to get SNN controllers is by converting the existing ANN controllers to SNNs ensuring the desired accuracy level. However, the size of the temporal window of an SNN plays a crucial role in the performance of SNNs. With a larger temporal window, an SNN can achieve better accuracy but at the cost of computational resources and latency. Moreover, the SNN controllers need to satisfy the underlying safe range specification to be considered as a promising alternative to the ANN counterparts. In this paper, we address the problem of determining the appropriate temporal window of an SNN controller such that the SNN is both accurate w.r.t. its respective ANN controller and ensures the safe range specification as well through formal verification. We have experimented on five benchmark neural controllers. For some of the SNN controllers that violate their safe range, we provide an iterative bound tightening method to approximate the safe range of the SNN controllers. Our method is expected to be highly valuable for safety verification and output range estimation of SNNs, with their growing popularity, mainly attributed to their low energy requirements. The ability to translate to SNNs with the SRLA from ANNs, as well as to train them independently, will position the SRLA as a viable choice among SNN activations. 
 
 Developing a closed-loop implementation integrating both the plant and the SNN controller is the immediate future step of this work. Moreover, while we currently focus on SNN controllers with SRLA, in the future, we aim to extend this work to support other SNN neuron models and neural controllers with other activations like sigmoid and tanh. Exploring more real-world benchmarks and doing a comprehensive comparison with existing works including ANN controllers can be done as another future step. In another direction, the effectiveness of the random simulation steps in our algorithms points to an exploration of directed fast falsification methods~\cite{Yu_Qian_Hu_2016} in conjunction with formal verification for better performance.

%% file: main.bbl
\begin{thebibliography}{10}

\bibitem{Mnih2015HumanlevelCT}
Volodymyr~Mnih et~al.
\newblock Human-level control through deep reinforcement learning.
\newblock {\em Nature}, 518:529--533, 2015.

\bibitem{Pavithra}
Pavithra Prabhakar and Zahra Rahimi~Afzal.
\newblock Abstraction based output range analysis for neural networks.
\newblock In {\em NEURIPS}, pages 15762--15772, 2019.

\bibitem{polar}
Chao Huang, Jiameng Fan, Xin Chen, Wenchao Li, and Qi~Zhu.
\newblock Polar: A polynomial arithmetic framework for verifying neural-network controlled systems.
\newblock In {\em International Symposium on Automated Technology for Verification and Analysis}, page 414–430, 2022.

\bibitem{Bojarski2016EndTE}
Mariusz Bojarski et~al.
\newblock End to end learning for self-driving cars.
\newblock {\em ArXiv}, abs/1604.07316, 2016.

\bibitem{Julian2018DeepNN}
Kyle~D. Julian, Mykel~J. Kochenderfer, and Michael~P. Owen.
\newblock Deep neural network compression for aircraft collision avoidance systems.
\newblock {\em Journal of Guidance, Control, and Dynamics}, 42(3):598--608, 2019.

\bibitem{Katz2017ReluplexAE}
Guy Katz, Clark Barrett, David~L. Dill, Kyle Julian, and Mykel~J. Kochenderfer.
\newblock Reluplex: An efficient smt solver for verifying deep neural networks.
\newblock In {\em Computer Aided Verification}, pages 97--117, Cham, 2017. Springer International Publishing.

\bibitem{main}
Elisabetta De~Maria, Cinzia Di~Giusto, and Laetitia Laversa.
\newblock Spiking neural networks modelled as timed automata with parameter learning, 2018.

\bibitem{vmcai}
Soham Banerjee et~al.
\newblock Smt-based modeling and verification of spiking neural networks: {A} case study.
\newblock In {\em VMCAI 2023}, pages 25--43, 2023.

\bibitem{partharoop2023}
Ankit Pradhan, Jonathan King, Srinivas Pinisetty, and Partha~S. Roop.
\newblock Model based verification of spiking neural networks in cyber physical systems.
\newblock {\em IEEE Transactions on Computers}, 72(9):2426--2439, 2023.

\bibitem{snnconvwang}
Bingsen Wang et~al.
\newblock A new ann-snn conversion method with high accuracy, low latency and good robustness.
\newblock In Edith Elkind, editor, {\em Proceedings of the Thirty-Second International Joint Conference on Artificial Intelligence 2023}, pages 3067--3075. International Joint Conferences on Artificial Intelligence Organization, 8 2023.
\newblock Main Track.

\bibitem{neuromorphic_computing}
Amar Shrestha, Haowen Fang, Zaidao Mei, Daniel~Patrick Rider, Qing Wu, and Qinru Qiu.
\newblock A survey on neuromorphic computing: Models and hardware.
\newblock {\em IEEE Circuits and Systems Magazine}, 22(2):6--35, 2022.

\bibitem{neuromorphic}
Zheqi Yu, Amir~M. Abdulghani, Adnan Zahid, Hadi Heidari, Muhammad~Ali Imran, and Qammer~H. Abbasi.
\newblock An overview of neuromorphic computing for artificial intelligence enabled hardware-based hopfield neural network.
\newblock {\em IEEE Access}, 8:67085--67099, 2020.

\bibitem{Bekolay2014}
Trevor Bekolay et~al.
\newblock Nengo: a {Python} tool for building large-scale functional brain models.
\newblock {\em Frontiers in Neuroinformatics}, 7(48):1--13, 2014.

\bibitem{asyncSN}
Amirreza Yousefzadeh~et al.
\newblock Asynchronous spiking neurons, the natural key to exploit temporal sparsity.
\newblock {\em IEEE Journal on Emerging and Selected Topics in Circuits and Systems}, 9(4):668--678, 2019.

\bibitem{dutta2017output}
Souradeep Dutta, Susmit Jha, Sriram Sanakaranarayanan, and Ashish Tiwari.
\newblock Output range analysis for deep neural networks, 2017.

\bibitem{ARCH19}
Diego~Manzanas Lopez et~al.
\newblock Arch-comp19 category report: Artificial intelligence and neural network control systems for continuous and hybrid systems plants.
\newblock In {\em International Workshop on Applied Verification of Continuous and Hybrid Systems}, volume~61, pages 103--119, 2019.

\bibitem{ARCH22}
Gidon Ernst et~al.
\newblock Arch-comp 2022 category report: Falsification with ubounded resources.
\newblock In {\em International Workshop on Applied Verification of Continuous and Hybrid Systems}, volume~90, pages 204--221, 2022.

\bibitem{snnconvcao}
Yongqiang Cao, Yang Chen, and Deepak Khosla.
\newblock Spiking deep convolutional neural networks for energy-efficient object recognition.
\newblock {\em International Journal of Computer Vision}, 113(1):54--66, May 2015.

\bibitem{snnconvsengupta}
Abhronil Sengupta, Yuting Ye, Robert Wang, Chiao Liu, and Kaushik Roy.
\newblock Going deeper in spiking neural networks: Vgg and residual architectures.
\newblock {\em Frontiers in Neuroscience}, 13, 2019.

\bibitem{snnconvgao}
Haoran Gao et~al.
\newblock High-accuracy deep ann-to-snn conversion using quantization-aware training framework and calcium-gated bipolar leaky integrate and fire neuron.
\newblock {\em Frontiers in Neuroscience}, 17, 2023.

\bibitem{hoang_19}
Hoang-Dung Tran et~al.
\newblock Safety verification of cyber-physical systems with reinforcement learning control.
\newblock {\em ACM Transactions on Embedded Computing Systems}, 18(5s), oct 2019.

\bibitem{reachSherlock}
Souradeep Dutta, Xin Chen, and Sriram Sankaranarayanan.
\newblock Reachability analysis for neural feedback systems using regressive polynomial rule inference.
\newblock In {\em 22nd ACM International Conference on Hybrid Systems: Computation and Control ({HSCC})}, pages 157--168, 04 2019.

\bibitem{2020reachnn}
Jiameng Fan, Chao Huang, Xin Chen, Wenchao Li, and Qi~Zhu.
\newblock Reachnn*: A tool for reachability analysis of neural-network controlled systems.
\newblock In Dang~Van Hung and Oleg Sokolsky, editors, {\em Automated Technology for Verification and Analysis}, pages 537--542, Cham, 2020. Springer International Publishing.

\bibitem{verisig2}
Radoslav Ivanov, Taylor Carpenter, James Weimer, Rajeev Alur, George Pappas, and Insup Lee.
\newblock Verisig 2.0: Verification of neural network controllers using taylor model preconditioning.
\newblock In Alexandra Silva and K.~Rustan~M. Leino, editors, {\em Computer Aided Verification}, pages 249--262, Cham, 2021. Springer International Publishing.

\bibitem{gurobi}
{Gurobi Optimization, LLC}.
\newblock {Gurobi Optimizer Reference Manual}, 2023.

\bibitem{xiang2018verification}
Weiming Xiang, Patrick Musau, Ayana~A. Wild, Diego~Manzanas Lopez, Nathaniel Hamilton, Xiaodong Yang, Joel Rosenfeld, and Taylor~T. Johnson.
\newblock Verification for machine learning, autonomy, and neural networks survey, 2018.

\bibitem{Grossmann2002}
Ignacio~E. Grossmann.
\newblock Review of nonlinear mixed-integer and disjunctive programming techniques.
\newblock {\em Optimization and Engineering}, 3(3):227--252, Sep 2002.

\bibitem{cheng2017maximum}
Chih-Hong Cheng, Georg Nührenberg, and Harald Ruess.
\newblock Maximum resilience of artificial neural networks, 2017.

\bibitem{randbs}
Eitan Zemel.
\newblock Random binary search: A randomizing algorithm for global optimization in r1.
\newblock {\em Mathematics of Operations Research}, 11(4):651--662, 1986.

\bibitem{snnconvdiehl}
Peter~Udo Diehl, Daniel Neil, Jonathan Binas, Matthew Cook, Shih-Chii Liu, and Michael Pfeiffer.
\newblock Fast-classifying, high-accuracy spiking deep networks through weight and threshold balancing.
\newblock {\em 2015 International Joint Conference on Neural Networks (IJCNN)}, pages 1--8, 2015.

\bibitem{snnconvrueckauer}
Bodo Rueckauer, Iulia-Alexandra Lungu, Yuhuang Hu, Michael Pfeiffer, and Shih-Chii Liu.
\newblock Conversion of continuous-valued deep networks to efficient event-driven networks for image classification.
\newblock {\em Frontiers in Neuroscience}, 11, 2017.

\bibitem{snnconvding}
Jianhao Ding, Zhaofei Yu, Yonghong Tian, and Tiejun Huang.
\newblock Optimal ann-snn conversion for fast and accurate inference in deep spiking neural networks.
\newblock In Zhi-Hua Zhou, editor, {\em Proceedings of the Thirtieth International Joint Conference on Artificial Intelligence, {IJCAI-21}}, pages 2328--2336. International Joint Conferences on Artificial Intelligence Organization, 8 2021.
\newblock Main Track.

\bibitem{snnmv2018}
Elisabetta Maria, Cinzia Di~Giusto, and Laetitia Laversa.
\newblock Spiking neural networks modelled as timed automata with parameter learning.
\newblock {\em Natural Computing}, 19:135--155, 03 2020.

\bibitem{snnmv2016}
Bogdan Aman and Gabriel Ciobanu.
\newblock Modelling and verification of weighted spiking neural systems.
\newblock {\em Theoretical Computer Science}, 623(C):92–102, apr 2016.

\bibitem{AmirSK21}
Guy Amir, Michael Schapira, and Guy Katz.
\newblock Towards scalable verification of deep reinforcement learning.
\newblock In {\em Formal Methods in Computer Aided Design, {FMCAD} 2021, New Haven, CT, USA, October 19-22, 2021}, pages 193--203. {IEEE}, 2021.

\bibitem{Yu_Qian_Hu_2016}
Yang Yu, Hong Qian, and Yi-Qi Hu.
\newblock Derivative-free optimization via classification.
\newblock {\em Proceedings of the AAAI Conference on Artificial Intelligence}, 30(1), Mar. 2016.

\end{thebibliography}
